\documentclass[aps,prd,showkeys,nofootinbib,superscriptaddress]{revtex4-1}

\usepackage{enumerate}
\usepackage{amsmath,slashed,tensor,amssymb,epsfig,amsthm,bm,graphicx,graphics,slashed}
\usepackage[caption=false]{subfig}
\usepackage{hyperref}
\usepackage[utf8]{inputenc}
\usepackage[top=1.5cm, bottom=1.5cm, left=2cm, right=2cm]{geometry}

\usepackage{makecell}
\usepackage{microtype}
\usepackage[font=footnotesize,labelfont=bf]{caption}

\newcommand{\be}{\begin{equation}}\newcommand{\ee}{\end{equation}}
\newcommand{\bea}{\begin{eqnarray}}\newcommand{\eea}{\end{eqnarray}}
\newcommand{\brr}{\begin{array}}\newcommand{\err}{\end{array}}
\newcommand{\bit}{\begin{itemize}}\newcommand{\eit}{\end{itemize}}
\newcommand{\ben}{\begin{enumerate}}\newcommand{\een}{\end{enumerate}}

\newcommand{\bbm}{\begin{bmatrix}}\newcommand{\ebm}{\end{bmatrix}}
\newcommand{\ba}{\begin{array}}
\newcommand{\ea}{\end{array}}

\newtheorem{mydef}{Definition}
\newtheorem{Lemma}{Lemma}
\newtheorem{theorem}{Theorem}
\newcommand{\bd}{\begin{mydef}} \newcommand{\ed}{\end{mydef}}
\newcommand{\bthe}{\begin{theorem}} \newcommand{\ethe}{\end{theorem}}
\newcommand{\ble}{\begin{Lemma}} \newcommand{\ele}{\end{Lemma}}

\def\1{{_{1}}}\def\2{{_{2}}}

\def\noHe0{:\;\!\!\;\!\!:H_e(0):\;\!\!\;\!\!:}
\def\noHm0{:\;\!\!\;\!\!:H_\mu(0):\;\!\!\;\!\!:}

\def\1{{_{1}}}\def\2{{_{2}}}

\begin{document}

\title{Carbon pseudospheres and the BTZ black hole}

\author{A.~Iorio}
\email{iorio@ipnp.troja.mff.cuni.cz}

\affiliation{Institute of Particle and Nuclear Physics, Faculty  of  Mathematics  and  Physics, Charles  University, V  Hole\v{s}ovi\v{c}k\'{a}ch  2, 18000  Praha  8,  Czech  Republic.}


\begin{abstract}
I first recall the uses of Dirac materials as table top realizations of high energy physics scenarios. Then I point to a specific system that might reproduce a massless BTZ black hole, where the key role is played by hyperbolic carbon pseudospheres. Finally some considerations are offered on the possibility to realize rotating black holes, along with some comments on the future of the whole analog gravity enterprise.
\end{abstract}

\vspace{-1mm}

\maketitle

\section{Introduction}

The field of analogs has as noble father Richard Feynman who ignited the field in a famous lecture titled ``Electrostatic Analogs'', available in \cite{Feynman} (see also \cite{twostories}). There he explains why analog systems do describe the same physics, based on the thrilling hypothesis of \textit{more elementary constituents} than the ones we deem to be fundamental. Amazingly, when space itself is included as an emergent phenomenon, these are also the conclusions of certain completely independent arguments of contemporary quantum gravity \cite{bekenstein,scholtz,carroll}.

As for the field of \textit{gravity} analogs, see \cite{Volovik:2003fe}, the seminal paper is that of Unruh of 1981, where he proposes to search for experimental signatures of his and Hawking's effects, in a fluid dynamical analog~\cite{UnruhAnalog}. Due to our deeper understanding and experimental control of condensed matter systems, it is now becoming increasingly popular to reproduce that and other aspects of fundamental physics in analog systems. Examples include the Hawking phenomenon in Bose--Einstein condensates~\cite{Steinhauer:2015saa}, the Weyl symmetry \cite{iorio} and the related Hawking/Unruh phenomenon on graphene~\cite{ioriolambiase1}, gravitational and axial anomalies in Weyl semimetals~\cite{Gooth:2017mbd}, and more~\cite{Ulf_LeonhardtPRL2019}. Actually, gravity analogs are not limited to condensed-matter systems, as can be seen, e.g., by interpreting hadronization in heavy-ion collisions as a consequence of the Unruh effect~\cite{Castorina:2007eb,Castorina:2008gf}.

Despite those impressive advances, there are still two milestones to reach. One is to understand the epistemic role of analogs in fundamental hugh energy physics, as not all theorists would agree that analogs are much more than mere divertissements. In fact, experimental results obtained in analogs are not used as feedbacks for the target theories they are analogs of (see, e.g.,~\cite{Dardashti2016,twostories}). Anothe milestone would be a reliable definition of an analog BH entropy, or at least, of a QFT-like entanglement entropy that, in the presence of horizons, might serve the scope of setting-up some form of the second principle of BH thermodynamics.

Any progress in this direction would be truly important for the hep-th research. Having some results there, we could eventually be able to address the so-called {\it information paradox}, i.e., the apparent loss of information during BH evaporation, a question that, most probably, cannot be entirely solved via theoretical reasonings. See, e.g., \cite{Penrose1996,Hawking2004,Almheiri2013,Hooft2016,Hooft2016_I,Maldazena} for different points of view.

On the analog side, theoretical work has shown over the years that black hole physics can find an indirect realization in BEC systems~\cite{tris}, and thrilling experimental evidences have confirmed this fact~\cite{Steinhauer:2015saa}. The latter findings are often referred to as the first experimental examples of the Hawking effect. Here we focus on the proposal of graphene as an analog of high-energy fundamental physics\footnote{Inspired by those findings, fundamental constituents of both matter and space have been proposed in \cite{scholtz, twostories, smaldone}.}~\cite{ioriolambiase1,iorio,pabloStran,ioriopaiswitten,grapheneQFTreview,reach the unreachable}, based on the fact that its low-energy excitations~\cite{CastroNeto2009} are massless Dirac pseudo-relativistic fermions (the matter fields $\psi$), propagating in a carbon two-dimensional honeycomb lattice. The emergent (long-wave limit) description of the latter is a surface (piece of spacetime described by the ``emergent'' metric  $g_{\mu \nu}$). Such a behavior is shared by a wide range of materials, ranging from silicene and germanene through d-wave superconductors to topological insulators~\cite{wehling}. Each of those materials has its own peculiarities, which allow for further extensions of results obtained with graphene, and hence permit to explore a wider range of the high-energy target systems. Let us now give some details.

\section{Analog gravity on graphene}

Graphene is a one-atom-thick allotrope of carbon, i.e. the closest in nature to a 2-dimensional object. It was first theoretically speculated about \cite{wallace}, and, decades later, experimentally found \cite{geimnovoselovFIRST}. Its honeycomb lattice is made of two intertwined triangular sub-lattices. As is by now well known, this structure is behind the description of its electronic properties in terms of massless, (2+1)-dimensional, Dirac quasi-particles. If one linearizes the tight-binding Hamiltonian around two Fermi \textit{points},
$ \vec{k}^D_\pm = \left( \pm \frac{4 \pi}{3 \sqrt{3} \ell}, 0 \right)$, then the Hamiltonian becomes $H|_{\vec{k}_\pm} \simeq  v_F \sum_{\vec{p}} \left(\psi_+^\dagger \vec{\sigma} \cdot \vec{p} \; \psi_+  - \psi_-^\dagger \vec{\sigma}^* \cdot \vec{p} \; \psi_- \right)$,  where $v_F = 3 \eta \ell / 2 \sim c/300$ is the Fermi velocity, $\psi_\pm$ are two--component Dirac spinors, and $\vec{\sigma} \equiv (\sigma_1, \sigma_2)$, $\vec{\sigma}^* \equiv (\sigma_1, - \sigma_2)$, with $\sigma_i$ the Pauli matrices.

If one considers the linear regime only, the first scale is $E_\ell \sim v_F / \ell \sim 4.2$eV. Notice that $E_\ell \sim 1.5 \eta$, and that the associated wavelength, $\lambda = 2 \pi / |\vec{p}| \simeq 2 \pi v_F / E$, is $2 \pi \ell$. The electrons' wavelength, at energies below $E_\ell$, is large compared to the lattice length, $\lambda > 2 \pi \ell$. Those electrons see the graphene sheet as a continuum. One Dirac point is enough, when only strain is present (see, e.g., \cite{pabloStran}), and when certain approximations on the curvature are valid \cite{ioriolambiase1}. The importance and relevance of the two Dirac points for emergent hep-th descriptions has been discussed at length in our work \cite{ioriopaiswitten}, see also our recent \cite{tloop}, where the focus though is on torsion.

When only one Dirac point is necessary, the following Hamiltonian well captures the physics of undeformed (planar and unstrained) graphene:
$H = - i v_F \int d^2 x \; \psi^\dagger \vec{\sigma} \cdot \vec{\partial} \; \psi$, where the two component spinor is, e.g., $\psi \equiv \psi_+$, we moved back to configuration space, $\vec{p} \to - i \vec{\partial}$, and sums turned into integrals because of the continuum limit. In various papers, we have exploited this regime to a great extent, till the inclusion of curvature and torsion in the geometric background. On the other hand, we also have investigated the regimes beyond the linear one, where granular effects associated to the lattice structure emerge, see \cite{GUP} and also the related \cite{GUPBTZ}. When both Dirac points are necessary, one needs to consider four component spinors
$\Psi \equiv \left( \begin{array}{c} \psi_+ \\ \psi_- \\ \end{array} \right)$, and $4 \times 4$ Dirac matrices $\alpha^i = \left(\begin{array}{cc} \sigma^i & 0 \\ 0 & - {\sigma^*}^i \\ \end{array} \right)$, $\beta = \left(\begin{array}{cc} \sigma^3 & 0 \\ 0 & \sigma^3 \\ \end{array} \right)$, $i = 1, 2$. These matrices satisfy all the standard properties, see, e.g., \cite{grapheneQFTreview} and \cite{ioriopaiswitten}. With these, the Hamiltonian is $H  =  - i v_F \int d^2 x \left( \psi_+^\dagger \vec{\sigma} \cdot \vec{\partial} \; \psi_+
    - \psi_-^\dagger \vec{\sigma}^* \cdot \vec{\partial} \; \psi_- \right) =
    - i v_F \int d^2 x \; \bar{\Psi} \vec{\gamma} \cdot \vec{\partial} \; \Psi$.

In \cite{iorio} the goal was to identify the conditions for which graphene might realize aspects of QFT in curved spacetime. Therefore, key issues had to be faced, such as the proper inclusion of the time variable in a relativistic-like description, and the role of the nontrivial vacua and their relation to different quantization schemes for different observers. All of this finds its synthesis in the Unruh or the Hawking effects \cite{ioriolambiase1}. Let us explain here the main issues and the approximations made there.

Besides $E_\ell$, when we introduce curvature, we also have a second scale. When this happens, $E_\ell$ is our ``high energy regime''. This is so because we ask the curvature to be small compared to a limiting maximal curvature, $1/\ell^2$, otherwise: i) it would make no sense to consider a smooth metric, and ii) $r < \ell$ (where $1/r^2$ measures the intrinsic curvature), means that we should bend the very strong $\sigma$-bonds, an instance that does not occur. Therefore, our second scale is
$E_r \sim v_F / r$, with $E_r =  \ell / r \; E_\ell  < E_\ell$. To have a quantitative handle on these scales, let us take, e.g., $r \simeq 10 \ell$ as a small radius of curvature (high intrinsic curvature). To this corresponds an energy $E_r \sim 0.4$eV, whereas, to $r \sim 1 {\rm mm} \sim 10^6 \ell$, corresponds $E_r \sim 0.6 \mu$eV. The ``high energy'' to compare with is $E_\ell \sim 4$eV. When energies are within $E_r$ (wavelengths comparable to $2 \pi r$) the electrons experience the global effects of curvature. That is to say that, at those wavelengths, they can distinguish between a flat and a curved surface, and between, e.g., a sphere and a pseudosphere. Therefore, whichever curvature $r > \ell$ we consider, the effects of curvature are felt until the wavelength becomes comparable to $2 \pi \ell$. The formalism we have used, though, takes into account all deformations of the geometric kind, with the exception of torsion. Hence, this includes intrinsic curvature, and elastic strain of the membrane (on the latter see \cite{pabloStran}), but our predicting power stops before $E_\ell$, because there local effects (such as the actual structure of the defects) play a role that must be taken into account into a QG type of theory. On the latter the first steps were moved in \cite{GUP} (see also the related \cite{GUPBTZ}).

The intrinsic curvature is taken here as produced by disclination defects, that are customarily described in elasticity theory (see, e.g., \cite{Kleinert}), by the (smooth) derivative of the (non-continuous) SO(2)-valued rotational angle $\partial_i {\omega} \equiv {\omega_i}$, where $i=1,2$ is a curved spatial index. The corresponding (spatial) Riemann curvature tensor is easily obtained, ${R^{i j}}_{k l} =
    \epsilon^{i j} \epsilon_{k l} \epsilon^{m n} \partial_{m} \omega_{n} =
    \epsilon^{i j} \epsilon_{l k} 2 {\cal K}$,
where $\cal K$ is the Gaussian (intrinsic) curvature of the surface. In our approach we have included time, although the metric we adopted is
\begin{equation}\label{mainmetric}
g^{\rm graphene}_{\mu \nu}  = \left(\begin{array}{cc} 1 & 0  \quad 0 \\ \begin{array}{c} 0 \\ 0 \end{array} & g_{i j} \\ \end{array} \right)\;,
\end{equation}
i.e., the curvature is all in the spatial part, and $\partial_t g_{i j}= 0$. Since the time dimension is included, the SO(2)-valued (abelian) disclination field has to be lifted-up to a SO(1,2)-valued (non-abelian) disclination field, ${\omega_\mu}^a$, $a=0,1,2$, with $\omega_\mu^{\; a} = e^b_\mu \omega_b^{\; a}$ and the expression $\omega_a^{\; d}  = \frac{1}{2} \epsilon^{b c d} \left( e_{\mu a} \partial_b E_c^\mu + e_{\mu b} \partial_a E_c^\mu + e_{\mu c} \partial_b E_a^\mu \right)$, gives the relation between the disclination field and the metric (dreibein). All the information about intrinsic curvature does not change. For instance, the Riemann curvature tensor, ${R^\lambda}_{\mu \nu \rho}$, has only one independent component, proportional to $\cal K$ (see \cite{iorio}). When only curvature is important, the long wavelength/small energy electronic properties of graphene, are well described by the action ${\cal A} = i  v_F \int d^3 x \sqrt{g} \; \bar{\Psi} \gamma^\mu (\partial_\mu + \Omega_\mu) \Psi$, with $\Omega_\mu \equiv {\omega_\mu}^a J_a$, and $J_a$ are the generators of SO(1,2), the local Lorentz transformations in this lower-dimensional setting. In \cite{ioriopaiswitten} we have discussed at length this action within the Witten approach \cite{witten3dgravity} to Poincar\'{e} ($ISO(2,1)$) or (A)dS gravity as gauge theory, and within the USUSY approach, see also \cite{susyZanelli1}, and especially the recent \cite{u-susy-graphene}.

\begin{figure}[tbp]
\begin{center}
\includegraphics[width=0.3\textwidth]{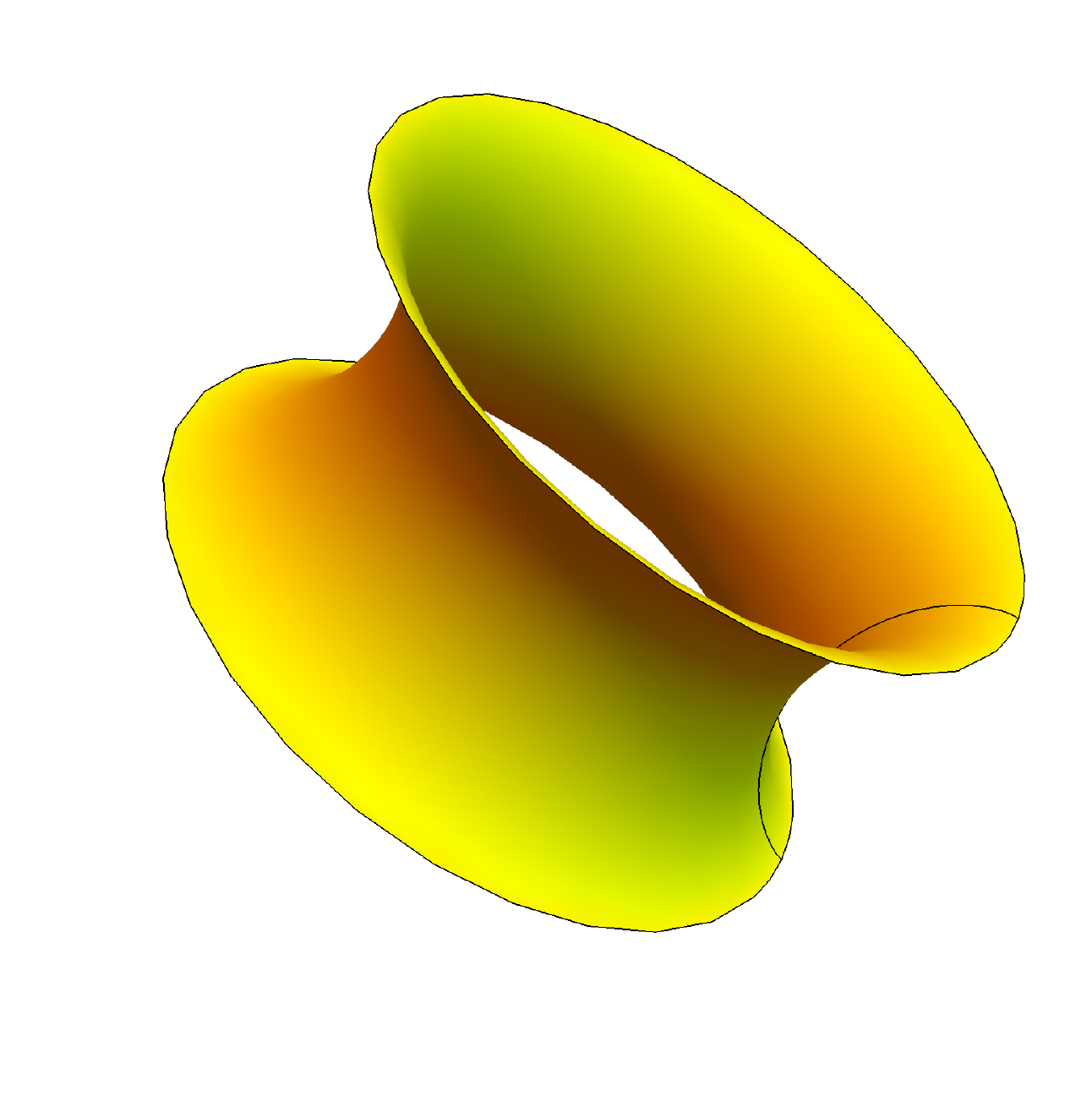}
\end{center}
\caption{\label{fig1} The hyperbolic pseudosphere for $a=1$, $C=1$. Here $\rho_{\rm min}= 1$ and $\rho_{\rm max} \simeq 1.4142$.} \label{hyperbolicwormhole1}
\end{figure}

Within this approach, a nontrivial $g_{tt}$ in (\ref{mainmetric}), hence a clean nontrivial general relativistic effect (recall that $g_{tt} \sim U_{grav}$) can only happen if specific symmetries and set-ups map the lab system into the wanted one, one being the local Weyl symmetry. This produced a measurable predictions of a Hawking/Unruh effect, for certain specific shapes. It was ofund that surfaces of constant Gaussian curvature, and among them, those of negative curvature (that necessarily have singular boundaries, see \cite{grapheneQFTreview} and \cite{icrystals}) are key. The above lead to the proposal of a variety of set-ups, especially three key spacetimes with horizon: the Rindler, the de Sitter and the BTZ black hole \cite{BTZ1992}.

\section{Carbon pseudospheres and the BTZ black hole}

Let us write the BTZ black hole metric as in \cite{GUPBTZ}
\be\label{btzgeneral}
ds_{BTZ}^2 = f(r)^{2} c^2dt^2 - f(r)^{-2}dr^2 - r^2(d\phi + N^\phi cdt)^2
\ee
where
\be
f^2(r) = -\frac{8GM}{c^2} - \Lambda r^2 + \frac{16G^2J^2}{c^4\,r^2}\,, \quad \quad \quad
N^\phi=-\frac{4GJ}{c^2\,r^2}\,,
\label{btzcarlip}
\ee
with $M$ the mass, $\Lambda \equiv - 1/\ell^2 < 0$ the negative cosmological constant and $J$ the angular momentum. Horizons are located at the positive zeros of $f(r)$
\be\label{horizonbtz}
r^2_\pm = \frac{4GM \ell^2}{c^2}\left[1 \pm \left(1 - \frac{J^2}{\ell^2 M^2} \right)^{1/2}\right] \,.
\ee
When $M> 0$ and $|J| \leq M \ell$, we have a black hole, with $r_+$ a genuine event horizon, and $r_-$ a Cauchy horizon.

\subsection{The massless black hole}

\begin{figure}[tbp]
\begin{center}
\includegraphics[width=0.3\textwidth]{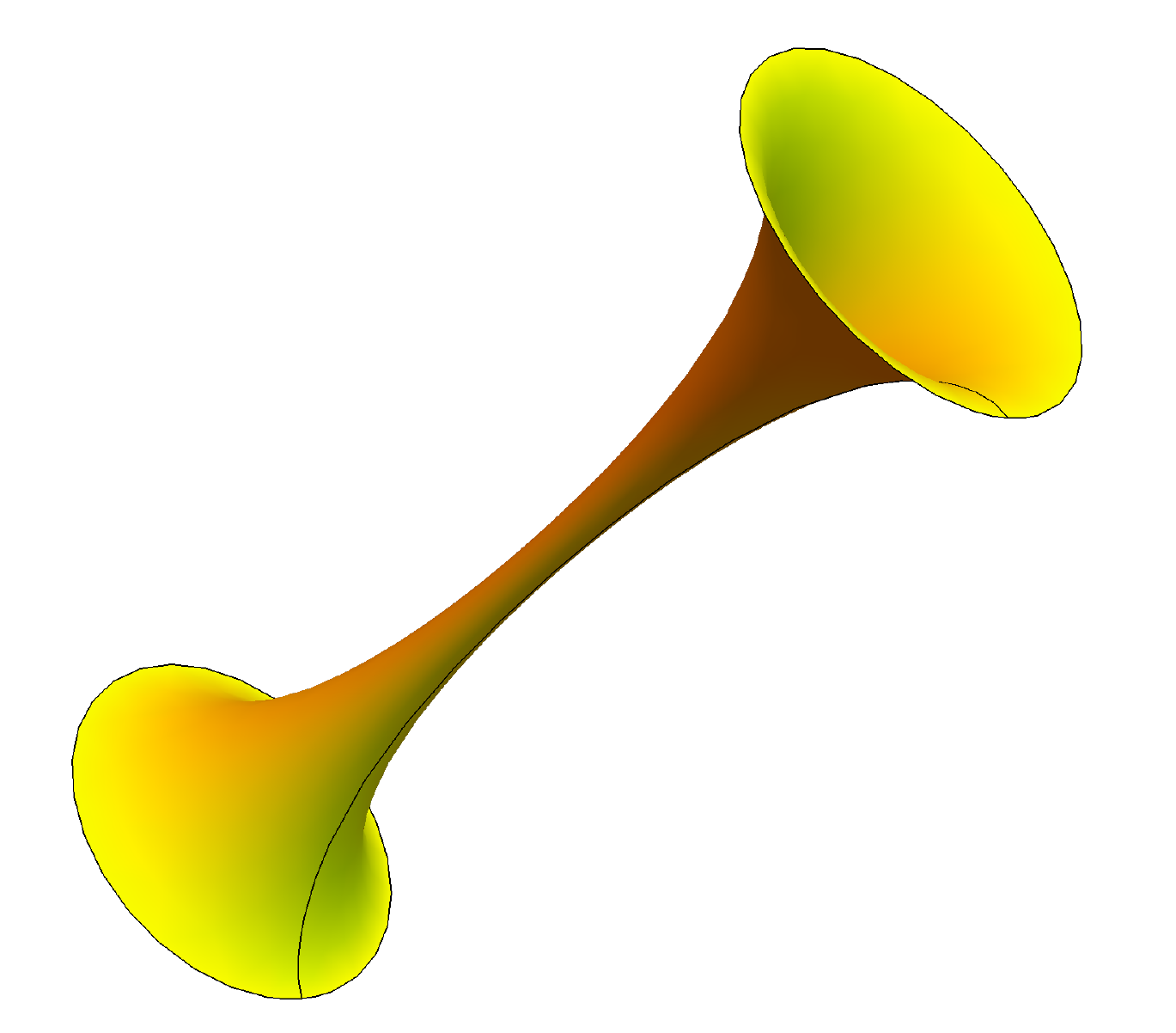}
\end{center}
\caption{\label{fig2} The hyperbolic pseudosphere for $a=1$, $C=1/10$. Here $\rho_{\rm min}= C = 1/10$ and $\rho_{\rm max} \simeq 1.005$.} \label{hyperbolicwormhole2}
\end{figure}

\begin{figure}[tbp]
\begin{center}
\includegraphics[width=0.7\textwidth]{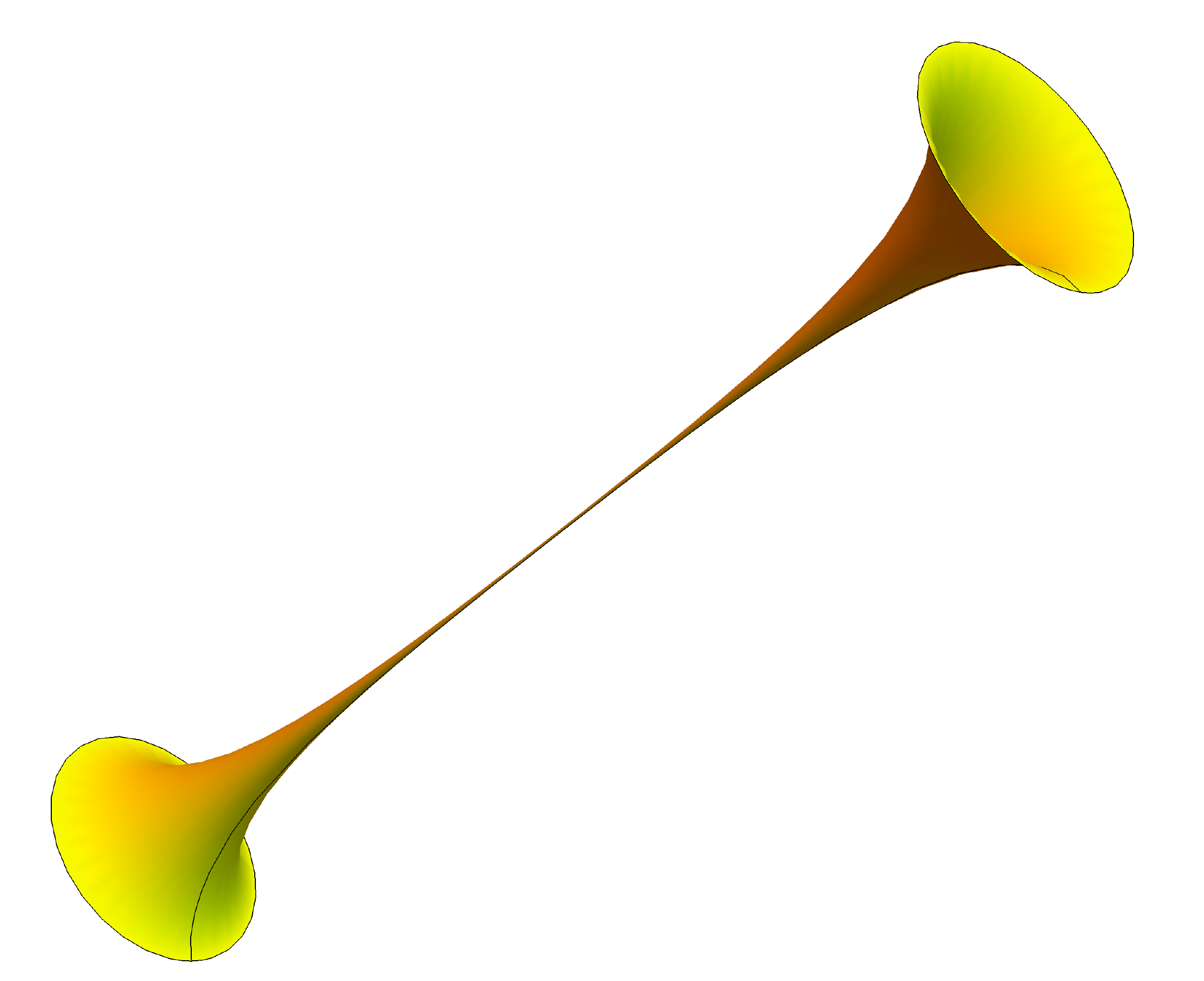}
\end{center}
\caption{\label{fig3} The hyperbolic pseudosphere for $a=1$, $C=1/100$. Here $\rho_{\rm min}= C = 1/100$ and $\rho_{\rm max} \simeq 1.00005$.} \label{hyperbolicwormhole3}
\end{figure}

In \cite{GUPBTZ}, see also \cite{ioriolambiase1}, the extremal case of $M \to 0$
\be \label{zerobtz}
ds_0^2 = (r/\ell)^{2} c^2 dt^2 - (r/\ell)^{-2}dr^2 - r^2 d\phi^2 \;,
\ee
was put into (conformal) correspondence to a graphene analog realization of this scenario, when shaped in a very specific manner. There it is shown that the most natural choice is to identify $\ell$ with $\ell_L$, the lattice spacing. The shape that is necessary to realize is that of the hyperbolic pseudosphere $\Sigma_{\rm HYP}$, see the figures, whose line element is
\be
dl_{\rm HYP}^2 = du^2 + C^2 \cosh^2(u/a) d\phi^2 \,,
\ee
with $C = \ell_L$, $u$ the longitudinal coordinate, $\phi \in [0,2 \pi]$ and the constant negative Gaussian curvature given by $K = - 1/a^2 < 0$. In terms of the radial coordinate
\be \label{radiusHyp}
\rho(u) = C \cosh(u/a) \,,
\ee
there is a singular boundary at the largest circle of radius $\rho_{\rm max} = \sqrt{a^2 + C^2} \equiv \rho_{Hh}$, where $C$ is the radius of the smallest throat, $\rho_{\rm min} = C$, cf. Figs.\ref{fig1}, \ref{fig2} and \ref{fig3}.

Indeed, setting $J=0$, and easing a little the notation, $8 G / c^2 = 1$
\be\label{btzhyp}
ds^2_{BTZ} = \left(r^2/C^2 - M \right) ds^2_{\rm HYP} \;,
\ee
where $ds^2_{\rm HYP}$ is the line element of $\Sigma_{\rm HYP} \times \mathbf{R}$. Therefore, all the relevant quantities of the BTZ black hole are given in terms of measurable quantities
\be\label{identifBTZ}
\Lambda \equiv - 1 / \ell_L^2 \quad , \quad M \equiv \ell_L^2 / a^2 \quad , \quad r_+ \equiv \ell_L^2 / a \;.
\ee

Before moving to the discussion of the $J\neq0$ case, let us recall how the event horizon, $r_+$ relates to the singular boundary of the hyperbolic pseudosphere \cite{ioriolambiase1,GUPBTZ}
\be
r_{H h} \equiv r (u_{H h}) = r_+ \coth\left( {\rm arccosh} \left( \sqrt{1 + a^2/\ell_L^2} \right) \right) \;.
\ee
In the limit of small $\ell_L/a$, these two horizons coincide. That is also the limit where $M \to 0$, and, accordingly $r_+ \to 0$, i.e. the announced zero mass black hole. In the figures we show three cases of such pseudosphere, for increasing closeness of $r_{Hh}$ to $r_+$.

There are three reasons to still call this a black hole.

First of all, it is continuously connected to the spectrum of BTZ black holes, $M \ge 0$. In fact, unlike the $M < 0$ cases, which are states under the black hole threshold, and represent point-particles in $AdS_3$, the $M=0$ case is not a point-particle. One way of looking at it is that in the flat limit $\Lambda \to 0$ (i.e. $\ell \to \infty$), where point particles survive, the case $M=0$ does not exist as a solution. Thus $M=0$ belongs more to the black holes segment/set ($M>0$) than to the point-particles ($M<0$). On the point particles see \cite{DJtH}.

Second, the metric (\ref{zerobtz}) has a horizon, even though it is not an event horizon. Indeed, the massless BTZ is locally equivalent to $AdS_3$ in Poincar\'{e} coordinates. If one decompactifies the angular coordinate, one finds that the BTZ metric with $M=0$ is the $AdS_3$ Poincar\'{e} patch, since the ``Poincar\'{e} horizon'' coincides with $r=0$.
The Poincar\'{e} horizon is the $r=0$ of the $M=0$ BTZ in BTZ coordinates. It is written in coordinates $z=1/r$ sometimes. In that case, obviously, the Poincar\'{e} horizon is $z=\infty$ while the boundary of the space is $z=0$. It is well known, so it is not a result per se, but have a look, e.g., at \cite{braga}.

A third argument (related to the first) in favor of calling it a black hole is that, from the point of view of holography, this solution ($M=0$) corresponds to a well-defined state of the dual theory (although it is a state of $T=0$). About the states of the dual theory, one should consider the link between three-dimensional gravity and two-dimensional Liouville field theory, see, e.g. the nice review \cite{modave}, and also \cite{wittenM=0}, where the ``black hole threshold'' is precisely the gap between $AdS_3$ and the $M=0$ BTZ.

\subsection{The rotating black hole?}

It is an intriguing and challenging question whether we could reproduce with graphene a \textit{rotating} black hole, $J \neq 0$, that is whether we could find a specific configuration that can be associated to the spacetime described by (\ref{btzgeneral}). One crucial observation is that such configuration needs to include a nontrivial behavior along the time direction, on top of a nontrivial behavior along the space direction. Of course, we are referring here to the important role played in that metric by
\be \label{Jdef}
g_{t \phi} = 2 r^2 N^\phi = - J \,.
\ee

We have, in fact, at least two choices. One choice would be to insist with the strategy above illustrated, and look for a configuration that is conformal to the full BTZ metric, rather than conformal only to the $J=0$ case, see (\ref{btzhyp}). In this case, one should probably give up the possibility of having both, the inner and outer horizons, $r_\pm$, in favor of the most important inner horizon only. Given that now
\be
r_\pm = \frac{1}{2} \frac{\ell_L^4}{a^2} \left( 1 \pm (1 - a^2/\ell_L^4 J^2)^{1/2} \right) \,,
\ee
we loose spherical symmetry, hence we should expect a deformed hyperbolic pseudosphere. In that case, the parameter $J$ will have to be related to such a geometric deformation, that might as well be a static one. Therefore, one should not expect, in principle, $J$ to be a true angular momentum, like the one stemming from a spinning pseudosphere, but $J$ can rather be just a geometric measure of the deviation of actual boundary from the singular circle of the previous discussion, of ray $r_{Hh}$. At this point, though the road becomes quite steep. In fact, such a deformed surface might serve well the scope of visually reproducing the deformed horizons, at least the inner one, but then would probably fail to be a surface of constant Gaussian curvature. In that case, the strategy adopted in the previous discussion cannot be applied, because we would not be guaranteed to be in a conformally flat space time \cite{iorio}.

Therefore, why not trying a completely different road, that is to actually act on the time components of (\ref{mainmetric}), hence to construct the wanted metric directly? For this to work, we could shape the membrane along the radial direction in such a way to reproduce $1/f^2(r)$. This alone would clearly produce a flat metric, as the curvature would only be in one direction.  At this point it is crucial to remember that the metric we are discussing, is that experienced by the conductivity electron of the material, henceforth, at least in principle, this could be achieved by letting such electrons interact with a suitably fine-tuned external electromagnetic field \cite{PabloLaser}, in such a way that $U_{em}$ mimics the $U_{grav}$ that is in $g_{tt}$. Therefore, we need to act with a space-dependent external electromagnetic field in such a way that the $g_{tt}$ component is equal to $f^2(r)$, and we are \textit{nearly} done.

In fact, the question left to answer is: what about the reproduction of the crucial $g_{t \phi}$? This would probably be the most delicate step, because we need here a feedback between spatial action and temporal action. One way to go could be to use the electromagnetic field to \textit{induce} elastic strain of the membrane. If that happens, such a strain could  produce a gauge field $A_\mu^{strain}$, as customary, see, e.g., \cite{pabloStran}, that can be summed up to $A_\mu^{em}$, to obtain an overall term that mixes space and time component, that is the spatial feedback on the metric obtained acting on the time components, and \textit{viceversa}. Notice that, while the interaction withe the external field need be very fine tuned, in order to reproduce exactly the wanted $f^2(r)$, the feedback on the strain can be quite generic as that component is actually defining the $J$, as seen in (\ref{Jdef}), hence can be arbitrary.

All the above is very fascinating, but need a thoroughly investigation of all details, theoretical and experimental, that are beyond the scope of this essay, and the focus on ongoing research \cite{PabloLaser}.

\section{Conclusions} The exciting and rapidly evolving field of analog gravity is facing a new era. The interest is shifting from the reproduction of the \textit{kinematical} aspects of the Hawking/Unruh phenomenon, that has reached a climax of precision and accuracy, to the realization of some form of \textit{dynamics}, a very challenging problem. The most important phenomenon to study is black-hole evaporation, the most prominent dynamical phenomenon of quantum gravity, with its plethora of open fundamental issues, such as the possibility that information is not preserved in this process, etc. Having recalled here how Dirac materials lend themselves to realize crucial aspects of black hole physics, we believe that the search for the realizations of such \textit{dynamical} will benefit from Dirac materials, such as graphene, that showed already to be a powerful and versatile analog of both quantum fields on curved spaces and quantum gravity.

\section*{Acknowledgements}
The author is indebted to Gaston Giribet and Jorge Zanelli for discussions on the special status of the massless BTZ black hole. He gladly acknowledges support from Charles University Research Center (UNCE/SCI/013) and from the grant SVV No. 260576.


\begin{thebibliography}{99}
\bibitem{Feynman} R. Feynman, et al., 2006 The Feynman Lectures on Physics (Pearson/Addison-Wesley).
\bibitem{twostories} A.~Iorio, J. Phys.: Conf. Series \textbf{1275} (2019) 012013 [arXiv:1902.07096].
\bibitem{bekenstein} J.~D.~Bekenstein, Phys. Rev. D \textbf{23} (1981) 287; Phys. Rev. E \textbf{89} (2014) 1; Sci. Am. \textbf{289} (2003) 58.
\bibitem{scholtz} G.~Acquaviva, {A.~Iorio}, M.~Scholtz, Ann. Phys. \textbf{387} (2017) 317.
\bibitem{carroll} N~.~Bao, S.~M. Carroll, A.~Singh, Internat. J. Modern Phys. D 26 (12) (2017) 1743013.
\bibitem{Volovik:2003fe} G. E. Volovik, The Universe in a helium droplet, Int. Ser. Monogr. Phys. {\bf 117}  (2006) 1; C.~Barcel{\'o}, S.~Liberati, M.~Visser, Liv. Rev. Rel. {\bf 14} (2011) 3.
\bibitem{UnruhAnalog} W.G.~Unruh, Phys. Rev. Lett. {\bf 46} (1981) 1351.
\bibitem{Steinhauer:2015saa} J. Steinhauer et al., Nature \textbf{569} (2019) 688; J. Steinhauer, Nature Phys. {\bf 12} (2016) 959.
\bibitem{iorio} A.~Iorio, Ann. Phys. {\bf 326} (2011) 1334.
\bibitem{ioriolambiase1} {A.~Iorio}, G.~Lambiase, Phys. Let. B {\bf 716} (2012) 334; Phys. Rev. D {\bf 90} (2014) 025006.
\bibitem{Gooth:2017mbd} J. Gooth, et al, Nature {\bf 547} (2017) 324.
\bibitem{Ulf_LeonhardtPRL2019} U. Leonhardt et al, Phys. Rev. Lett. {\bf 122} (2019) 010404.
\bibitem{Castorina:2007eb} P.~Castorina, D.~Kharzeev, H.~Satz, Eur. Phys. J. C {\bf 52} (2007) 187.
\bibitem{Castorina:2008gf} P.~Castorina, {A.~Iorio}, H.~Satz, Int. J. Mod. Phys. E {\bf 24} (2015) 1550056; P.~Castorina, D.~Grumiller, {A.~Iorio}, Phys. Rev. D {\bf 77} (2008) 124034.
\bibitem{Dardashti2016} R.~Dardashti, K.~P.~Thebault, E.~Winsberg, Brit. J. Phil. Science {\bf 68} (2015) 55.
\bibitem{Penrose1996} R.~Penrose, Gen. Rel. Grav. {\bf 28}  (1996) 581.
\bibitem{Hawking2004} S.~Hawking, in {\em General relativity and gravitation}, Proc. GR17, Dublin (2004) pp. 56.
\bibitem{Almheiri2013} J.~Polchinski, et al.,  J. High Energy Phys. {\bf 2013} (2013) 62.
\bibitem{Hooft2016}  R.~B.~Mann, {\em Black Holes: Thermodynamics, Information, and Firewalls} (Springer, Berlin, 2015).
\bibitem{Hooft2016_I} G.~'t Hooft, (2016), arXiv:1612.08640v1.
\bibitem{Maldazena} J.~Maldacena, L.~Susskind, Fortsch. Phys. {\bf 61} (2013) 781.
\bibitem{tris} R.~Balbinot, A.~Fabbri, S.~Fagnocchi, A.~Recati, I.~Carusotto, Phys. Rev. A {\bf 78} (2008) 021603(R).
\bibitem{smaldone} G.~Acquaviva, {A.~Iorio}, L.~Smaldone, Phys. Rev. D {\bf 102} (2020) 106002.
\bibitem{pabloStran} {A.~Iorio}, P.~Pais, Phys. Rev. D {\bf 92} (2015) 125005.
\bibitem{ioriopaiswitten} {A.~Iorio}, P.~Pais, Ann. Phys. {\bf 398} (2018) 265.
\bibitem{grapheneQFTreview} {A.~Iorio}, Int. J. Mod. Phys. D {\bf 24} 5 (2015) 1530013.
\bibitem{reach the unreachable} {A.~Iorio}, Frontiers in Materials {\bf 1} (2015) 36.
\bibitem{CastroNeto2009} A.~H.~Castro Neto, F.~Guinea, N.~M.~R.~Peres, K.~S.~Novoselov, A.~K.~Geim, Rev. Mod. Phys. {\bf 81} (2009) 109.
\bibitem{wehling} T.~O~Wehling, A.~M.~Black-Schaffer, A.~V.~Balatsky, Adv. Phys. {\bf 76} (2014) 1.
\bibitem{wallace} P.~R.~Wallace, Phys. Rev. {\bf 71} (1947) 622; G.~W.~Semenoff, Phys. Rev. Lett. {\bf 53} (1984) 2449.
\bibitem{geimnovoselovFIRST} K.~S.~Novoselov, A.~K.~Geim, et al, Science {\bf 306} (2004) 666.
\bibitem{tloop} M.~Ciappina, {A.~Iorio}, P.~Pais, A.~Zampeli, Phys. Rev. D \textbf{101} (2020) 036021.
\bibitem{GUP} A.~Iorio, P.~Pais, I.~A.~Elmashad, A.~F.~Ali, Mir Faizal, L.~I.~Abou-Salem, Int. J. Mod. Phys. D {\bf 27} (2018) 1850080.
\bibitem{GUPBTZ} A.~Iorio, G.~Lambiase, P.~Pais, F.~Scardigli, Phys. Rev. D \textbf{101} (2020) 105002.
\bibitem{Kleinert} H.~Kleinert, Gauge fields in condensed matter, Vol II, World Scientific (Singapore) 1989; M.~O.~Katanaev, I.~V.~Volovich, Ann. Phys. {\bf 216} (1992) 1.
\bibitem{witten3dgravity} E.~Witten, Nucl. Phys. B {\bf 311} (1988) 46.
\bibitem{susyZanelli1} J.~Zanelli, et al., J. High Energy Phys.  {\bf 1204} (2012) 058; J. High Energy Phys. {\bf 85} (2016) 201.
\bibitem{u-susy-graphene} L.~Andrianopoli, et al., J. High Energy Phys. \textbf{2001} (2020) 084.
\bibitem{icrystals} A.~Iorio, J. Phys.: Conf. Ser. \textbf{442} (2013) 012056; et al., J. Phys.: Cond. Matt. {\bf 28} (2016) 13LT01.
\bibitem{BTZ1992} M. Ba\~{n}ados, C. Teitelboim, J. Zanelli, Phys. Rev. Lett. 69 (1992) 1849.
\bibitem{entaentro} A.~Iorio, G.~Lambiase, G.~Vitiello, Ann. Phys. {\bf 309} (2004) 151.
\bibitem{gibbons} M.~Cvetic and G.~Gibbons, Ann. Phys. {\bf 327} (2012) 2617.
\bibitem{DJtH} S.~Deser, R.~Jackiw, G.~'t~Hooft, Three-dimensional Einstein gravity: Dynamics of flat space, Annals of Physics, {\bf 152} (1984) 220.
\bibitem{braga} C.~A.~Bayona and N.~R.~F.~Braga, Gen. Rel. Grav. {\bf 39} (2007) 1367.
\bibitem{modave} L.~Donnay, PoS {\bf Modave2015} (2016) 001 and references therein.
\bibitem{wittenM=0} J.-M.~Schlenker, E.~Witten, {\it No Ensemble Averaging Below the Black Hole Threshold}, arXiv:2202.01372 [hep-th].
\bibitem{PabloLaser} A.~Iorio, P.~Pais, {\it Laser-graphene interaction and the improvement of gravity analogs}, in preparation.
\end{thebibliography}

\end{document}